\documentclass[review,5p]{elsarticle}

\usepackage{hyperref}
\usepackage[latin1]{inputenc}
\usepackage{color}           
\usepackage{soul}            
\usepackage{amssymb}
\usepackage{amsfonts}
\usepackage{amsmath}
\usepackage{wasysym}
\usepackage{graphicx}
\usepackage{epsfig}
\usepackage{amsthm}
\usepackage{bm}

\journal{Journal of \LaTeX\ Templates}

\begin{document}

\begin{frontmatter}

\title{Universal quantum uncertainty relations:\\minimum-uncertainty wave packet depends on measure of spread}

\author{Anindita Bera\(^{1,2}\), Debmalya Das\(^{2}\), Aditi Sen(De)\(^{2}\), and Ujjwal Sen\(^{2}\)}
\address{\(^1\)Department of Applied Mathematics, University of Calcutta, 92 Acharya Prafulla Chandra Road,\\ Kolkata 700 009, India\\
\(^2\)Harish-Chandra Research Institute, HBNI,  Chhatnag Road, Jhunsi, Allahabad 211 019, India}




\begin{abstract}
Based on the statistical concept of 
the median, we propose a quantum uncertainty relation between  semi-interquartile ranges of the  position and momentum distributions of arbitrary quantum states. The relation is universal,  unlike that based on the mean and standard deviation, as the latter may become non-existent or  ineffective in certain cases. We show that the median-based one is not saturated for Gaussian distributions in position.  Instead, the Cauchy-Lorentz distributions in position turn out to be the one with the minimal uncertainty, among the states inspected, implying that the minimum-uncertainty state is not unique but depends on the measure of spread used. Even the ordering of the states with respect to the distance from the minimum uncertainty state is altered by a  change in the measure. We invoke the completeness of Hermite polynomials in the space of all quantum states  to probe the median-based relation. The results have potential applications in a variety of studies including those on the quantum-to-classical boundary and on quantum cryptography. 

\end{abstract}

\begin{keyword}
\texttt{Quantum uncertainty relation}\sep \texttt{Quantum information}\sep \texttt{Quantum mechanics}
\end{keyword}

\end{frontmatter}


\section{Introduction}
\label{intro}

The uncertainty relation is one of the most famous relations in quantum
mechanics. In its earliest and probably most well-known form, due to W. K.
Heisenberg and others \cite{gorom, kennard1, weyl}, it constrains the product of the
spreads of position and
momentum of an arbitrary quantum state to a certain lower limit. This lower limit 
is proportional to the Plank's constant, \(h\). This is certainly
not true in the classical mechanical description of the world, and it is
possible to know both the position and momentum of a single classical particle
with arbitrary precision. The uncertainty relation is a crucial element in multifarious  
branches of research, and  while it is important for interpretations of quantum mechanics including those 
related to the classical limit of quantum mechanics, it also has
far-reaching practical implications, for example in quantum cryptography \cite{Grudka-Michal-Ryszard}, and in gravitational wave detection \cite{ligo}.

It is natural to ask which quantum state 
provides the lowest product of the spreads of position and momentum. 
It is well-known that Gaussian
position distributions are minimum-uncertainty states
when considering the usual ``mean-based'' quantum uncertainty relation.
The ground state of a quantum harmonic oscillator and squeezed coherent states are examples 
of such minimal uncertainty states in physical systems.
Distributions that do not possess the mean and standard deviation, such as that of the Cauchy-Lorentz, 
are not considered in such searches for the minimal state. It may be noted that the notion of the onset of classicality or closeness to the classical situation can be conceptualized in several ways, one of which is the uncertainty relation, although there is at present no general consensus about which is the best \cite{ballentine94,dsen01,markham03,ballentine04,angelo05}.
Surprisingly, we find that 
the title of the  minimum-uncertainty states, as obtained via the uncertainty relation, is taken by a non-Gaussian state,  when we use the 
``median-based'' uncertainty relation, and the latter is 
a universal one, valid -- existing and efficient -- for all quantum states.

The quantification of the spread of a probability distribution
corresponding to a random variable typically begins with identifying a
figure of merit for the middle of the distribution -- a measure of
 ``central tendency'' -- and then the spread or measure of ``dispersion'' is
defined via that of central tendency. A popular measure of central
tendency of a distribution is the average of the values of the random
variable weighted by the corresponding probabilities. This is known as the
mean of the probability distribution. A measure of dispersion, called
standard deviation, is then obtained by considering the square root of the
mean of the squares of the deviations around the mean. 
It can for example be used to estimate the spread in the values obtained in a measurement.
As the epithet
implies, standard deviation happens to be a very useful, and consequently
widely-used, quantifier of the amount of variation of distributions ranging from climate science 
and financial services to medical science and sports
\cite{Feller-book, Amritava-Gupta-book}.

In quantum mechanics, it is therefore natural to express the Heisenberg uncertainty relation in
terms of standard deviations of the position and momentum
distributions of a quantum system, as was done in its original formulation \cite{gorom, kennard1}
and further extensions 
(see \cite{robertson, scrodinger-vai, schrodinger-translation, birula1, Deutsch, Partovi, new-birula, kraus, Maassen1, birula-dekh, 
Vicente,berta-nature, zozor, Branciard1, Branciard1a, Werner-uff,Pati-prl} and references therein). 
Enunciated in terms of
standard deviations, this ``mean-based'' uncertainty relation is saturated
by Gaussian position distributions. In other words, the product of the
standard deviations of position and momentum for all quantum states,
whose standard deviations do exist,
 is at
least a ``distance'' \(h\) (rather, proportional to \(h\)) away from what
is optimally possible in the classical world~\cite{Cohen-book, Bohm-book, parthasarathy-book}. In particular, Kramers~\cite{kramers-book} refers to  Gaussian states as the ``most favourable wave packets". And Baym~\cite{baym-book} refers to them as the ``most certain wave packets".

While mean is a pretty standard measure of central tendency, there are
other quantifications of the same that have their own merits. Some
well-known instances when the mean is not a measure of choice are (a) when
there are large outliers in a distribution, and (b) when the distribution
is skewed 
\cite{Feller-book, Amritava-Gupta-book}. 
Furthermore, (c) the mean of a
distribution may not exist, and this is of course another case where the
mean cannot be a choice. Quantum states can give rise to position and
momentum distributions that have some or all such features, and in such circumstances, the
mean-based uncertainty relations are either ineffective or non-existent.

The median of a probability distribution is the middle-most value of the
random variable when its values are arranged in a
certain order, say increasing. The median is unaffected by large outlier
values of the random variable, and provides a good estimate of central
tendency even for skewed distributions. Importantly, the median of a
distribution always exists. As a measure of dispersion, one can then use
the median of the moduli of the deviations around the median, a measure
referred to as the semi-interquartile range. Just like the median, the
semi-interquartile range exists for position and momentum distributions of
all quantum states.

The question that we address in this paper is whether quantum mechanics places a median-based uncertainty constraint on our knowledge of position and momentum
for arbitrary quantum states, that is independent of the mean-based one, and of wider applicability. We show by considering 
paradigmatic classes of position probability distributions corresponding to quantum states that such a constraint does exist, and has a 
rather different 
nature than the usual mean-based one. 
In particular, the minimum median-based uncertainty states are no more Gaussian distributions in position. 
The Cauchy-Lorentz distribution, due to M. G. Agnesi, S. D. Poisson,  A.-L. Cauchy, H. Lorentz, and others, provides the best choice among the distributions 
that we investigated, with the Gaussian being quite far off in the race. Therefore, while the Gaussian distribution in position 
saturates the 
mean-based uncertainty relation, 
the Cauchy-Lorentz distribution in position -- the square-root of the Cauchy-Lorentz distribution, up to a phase, being the corresponding 
quantum wave function in coordinate representation -- provides a lower uncertainty product from the perspective of the 
median-based uncertainty relation. The Cauchy-Lorentz distribution, being without a mean, was not even considered in the search for a saturating 
function of the mean-based uncertainty relation.
Even the ordering of states with respect to their distance to the minimum uncertainty state
is altered by the choice of measure of the spread of probability distributions.
We call upon the 
completeness of Hermite polynomials in \(L_2(-\infty, +\infty)\), the space of 
all possible states for a quantum system moving in one dimension,
to investigate the median-based relation.

\section{Setting the stage}
\label{stage}
Quantum mechanics asserts that there is a lower limit to the product of the standard deviations of physical quantities corresponding to 
``incompatible'' observables in any quantum state. 
We can consider incompatible observables to be those that do not share a common eigenstate. 
Suppose that  $A$ and $B$ represent
an arbitrary pair of incompatible 
observables.
For any quantum state $\rho$, the Cauchy-Schwartz inequality can be utilized to prove the 
mean-based uncertainty relation given by
\cite{sakurai-book, Ballentine-book}
\begin{equation}
  \Delta A \Delta B  \geq \frac{1}{2}
 \left|\left\langle \left[A,B\right]\right\rangle\right|,
 \label{gen_ur}
\end{equation}
where $\Delta A$ and $\Delta B$ are the uncertainties in the measurements
of the observables $A$ and $B$, respectively, in the state \(\rho\), as quantified by the corresponding 
standard deviations. The uncertainty $\Delta A$, in the state \(\rho\), is defined as 
\begin{equation}
 \Delta A = \sqrt{\langle (A  - \langle A \rangle)^2 \rangle},
 \label{u_def}
\end{equation}
and similarly for $\Delta B$. Equation~(\ref{u_def}) defines
the uncertainty in terms of the mean of the squares of deviations around the mean. For an arbitrary operator, \(\mathcal{A}\), 
the mean, \(\langle \mathcal{A} \rangle\), in the state \(\rho\), is defined by using the  
Born rule as 
\begin{equation}
\langle \mathcal{A} \rangle = \mbox{Tr}\left(\mathcal{A}\rho\right).
\end{equation}

While dealing with position ($x$) and momentum ($p$) measurements for an arbitrary quantum state, we note that
\begin{equation}
 \left[x,p\right]= i \hbar,
 \label{commutatorxp}
\end{equation}
where \(\hbar = h/(2\pi)\).
Thus the mean-based uncertainty relation for position and momentum reduces to
\begin{equation}
  \Delta x   \Delta p  \geq \frac{\hbar}{r},
\label{ur_xp-eiTa-purono}
\end{equation}
where \(r=2\).

As already mentioned, in spite of the fact that the mean of a probability distribution is used in an overwhelmingly large number of applications, 
there are definite cases in which the mean is not the most useful measure of central tendency. It is therefore useful to 
conceptualize measures of central tendency that are defined independent of the mean. One such is the median, which is the middle-most value 
of the random variable when the values of the random variable are arranged in a certain order, say increasing. More precisely, and 
focussing attention on continuous probability distributions, the median, \(\widetilde{\langle\mathcal{A}\rangle}\), 
of a probability distribution \(P(\mathcal{A}=a)\), corresponding to a 
random variable \(\mathcal{A}\), is obtained by solving the equation
\begin{equation}
 \int_{-\infty}^{\widetilde{\langle\mathcal{A}\rangle}} P(\mathcal{A}=a) da = \frac{1}{2}.
\end{equation}
In case the probability distribution is 
partly or fully discrete, the definition needs to be suitably changed. 

%
%
The concept of the median can however be used to arrive at definitions of dispersion that are independent of the standard deviation. 
An immediate set of measures of dispersion that depend on the median, can be obtained by considering the mean of the moduli of the deviations around the median, 
or the square root of the mean of the squares of the 
deviations around the median, etc. (cf. \cite{Werner-uff}). These however, in general, carry with them the troubles associated with the mean, as can e.g. be seen from the fact that the 
Cauchy-Lorentz distribution does possess these quantities as well. 
To see this, consider 
the Cauchy distribution, given by 
\begin{equation}
\label{bharsa-thakuk-aro-birol}
f_C(a:x_0,\gamma)=\frac{\gamma}{\pi} \frac{1}{(a-x_0)^2+\gamma^2},
\end{equation}
where \(a\), belonging to the range \((-\infty, \infty)\), represents the values of the corresponding random variable, and
 $\gamma >0$ as well as $x_0$ are the distribution parameters.
For the Cauchy distribution, the mean does not exist\footnote{The Cauchy principle value of the mean, however, does exist.}. But, as for all probability distributions, the Cauchy distribution does have a median. 
However, the mean of the moduli of the deviations around the median again does not exist. 
It is, therefore, worthwhile to 
set up a stage where the concept of the median can be utilized to 
provide a suitable measure of dispersion, that does not take recourse 
to the mean. One of the ways in which this can be attained is as follows. Just like the median that 
signals the point where the probability distribution is exactly
half-way, we can define ``quartiles'' that signal when the distribution is quarter-way and three-quarters-way. 
More precisely, 
the first and third quartiles, \(Q^{\mathcal{A}}_1\) and \(Q^{\mathcal{A}}_3\),
 of the probability distribution \(P(\mathcal{A}=a)\) are given by 
\begin{equation}
 \int_{-\infty}^{Q^{\mathcal{A}}_{2m+1}} P(\mathcal{A}=a) da = \frac{2m+1}{4},
\end{equation}
for \(m=0,1\).
The second quartile is of course the median itself. 
The ``semi-interquartile range'',
\begin{equation}
\widetilde{\Delta \mathcal{A}} = \frac{1}{2} \left( Q^{\mathcal{A}}_3 - Q^{\mathcal{A}}_1 \right),
 \end{equation}
is a suitable quantifier of dispersion based on the median, and we have manifestly steered clear of the concept of the mean. 

There exists an interesting ``width"-based single parameter family of measures of dispersion depending on a parameter
$\varepsilon>0$~\cite{Landau, Uffink, Busch}, that may be expected to have properties similar to the interquartile range,
for the case when $\varepsilon=\frac{1}{2}$. However the quantities are different, as can be seen, e.g., by considering the
second excited state of the one-dimensional quantum simple harmonic oscillator. Moreover, the uncertainty relations derived 
for the width-based quantities have no (non-zero) positive lower bound for $\varepsilon=\frac{1}{2}$~\cite{Landau, Uffink, Busch}.

For the time-independent pure quantum state, \(\psi(x)\), in coordinate representation, of a system moving on a straight line, 
the quartiles, \(Q^x_1\) and \(Q^x_3\), of the position probability distribution for that system can be obtained from
\begin{equation}
 \int_{-\infty}^{Q^{x}_{2m+1}} |\psi(x)|^2 dx = \frac{2m+1}{4},
\end{equation}
for \(m=0,1\).
For obtaining the quartiles, \(Q^p_1\) and \(Q^p_3\), corresponding to the momentum probability distribution of the same system, we may go over to the 
momentum representation of \(\psi(x)\), given by the Fourier transform,
\begin{equation}
 \phi(p) = \frac{1}{\sqrt{2 \pi \hbar}} \int_{-\infty}^\infty e^{-ixp/\hbar}\psi(x)dx.
\end{equation}
The quantities, \(Q^p_1\) and \(Q^p_3\), are then given by 
\begin{equation}
 \int_{-\infty}^{Q^{p}_{2m+1}} |\phi(p)|^2 dp = \frac{2m+1}{4},
\end{equation}
for \(m=0,1\).

\section{Median-based quantum uncertainty relation}
\label{median-uncertainty}

The quantity that we wish to analyze is the product of the semi-interquartile 
ranges of position and momentum for arbitrary quantum states.
Precisely, we consider 
\begin{equation}
 \widetilde{\Delta x} \widetilde{\Delta p},
\end{equation}
where 
\begin{equation}
 \widetilde{\Delta x} = \frac{1}{2} (Q^x_3 - Q^x_1), \quad  \widetilde{\Delta p} = \frac{1}{2} (Q^p_3 - Q^p_1).
\end{equation}
Like in relation (\ref{ur_xp-eiTa-purono}), our aim here is to find a lower bound of \(\widetilde{\Delta x} \widetilde{\Delta p}\), but for arbitrary quantum states.
We will now perform several case studies, in each of which we begin with a paradigmatic probability distribution for the position 
of a quantum system moving in one dimension.


\subsection{Cauchy-Lorentz distribution}
\label{cauchy}
The Cauchy-Lorentz probability distribution is  given 
in  equation~(\ref{bharsa-thakuk-aro-birol}).
Let us consider  a quantum system in one dimension whose wave function, $\psi_C(x)$, in coordinate representation, is given by 
\begin{equation}
\psi_C(x)=(f_C(x:x_0,\gamma))^{\frac{1}{2}},~ x \in \mathbb{R}.
\end{equation}
This function is square-integrable and continuous, and hence is a valid quantum mechanical wave function.
The wave function of course depends on the parameters \(x_0\) and \(\gamma\), which we have not included in the notation for the same.  
Since the mean of the Cauchy-Lorentz distribution is non-existent, the standard deviation necessarily does not exist. Therefore, for the 
system under consideration, the average position and the spread of the position distribution cannot be represented in terms of the 
mean and standard deviation of the same.  
We may however conveniently use the median and the semi-interquartile range for these purposes.

For position probability distribution function, 
$|\psi_C(x)|^2$, the values of first and third quartiles are $Q_{1,C}^x=x_0-\gamma$ and $Q_{3,C}^x=x_0+\gamma$. So, the semi-interquartile range 
$=\gamma$. For the momentum space wave 
function $\phi_C(p)$ corresponding to $\psi_C(x)$, the momentum probability distribution is given by
\begin{equation}
|\phi_C(p)|^2=\left|\frac{1}{\sqrt{2 \pi \hbar}} \int_{-\infty}^\infty e^{-ixp/\hbar}\psi_C(x)dx\right|^2.
\end{equation}
Now, to find the first quartile, $Q_{1,C}^p$, we need to solve the following equation:
\begin{equation}
\label{first-quartile-momentum}
\int_{-\infty}^{Q_{1,C}^p} |\phi_C(p)|^2 dp=\frac{1}{4}.
\end{equation}
We solve this equation numerically by 
considering the Gaussian quadrature and Van Wijngaarden-Dekker-Brent  methods \cite{tathastu}, to obtain the \(Q_{1,C}^p\) 
for $\gamma=1, 2, 3, 4$. 
The numerical values obtained are correct to three decimal places. 
In a similar way, we compute the third quartiles, $Q_{3,C}^p$.
for the same values of \(\gamma\).
The semi-interquartile ranges for $\phi_C(p)$ are given by $0.094 \hbar, 0.047 \hbar, 
0.032 \hbar, 0.024 \hbar$, for  $\gamma=1, 2, 3, 4$ respectively. 
The products of the semi-interquartile ranges, for $\gamma=1, 2, 3, 4$, are therefore 
\(\geq 0.094 \hbar$.
In other words, 
the median-based quantum uncertainty product, \(\widetilde{\Delta x} \widetilde{\Delta p}\), 
for 
the Cauchy-Lorentz position probability distribution, with the parameter \(\gamma = 1,2,3,4\), is 
\(\geq \frac{\hbar}{10.6}\).

Let us mention here that the Cauchy-Lorentz distribution has a finite differential entropy \cite{wehrl78} for $\gamma>0$. However, there exists quantum states for which the differential entropy diverges.  An example of such a state is $\sum_{k=2}^{\infty} \sqrt{p_k} |x=k\rangle$, where $p_k=\frac{1}{\alpha k (\ln k)^2}$, with $\alpha \approx 2.1$ \cite{wehrl78}. This implies that there exist states for which one cannot meaningfully consider the entropic quantum uncertainty relations \cite{birula1,Deutsch,Maassen1,Wehner12} although the median-based one does remain physically relevant and mathematically well-defined.

%
%


\subsection{Gaussian distribution}
\label{Gaussian}

The Gaussian distribution is a well-known continuous 
probability distribution, for which the probability distribution function is given by 
\begin{equation}
f_G(a:\mu, \sigma^2)=\frac{1}{\sqrt{2 \pi \sigma^2}} e^{-\frac{(a-\mu)^2}{2 \sigma^2}}, 
a \in \mathbb{R},
\end{equation}
where $\mu$ and \(\sigma^2\) are the distribution parameters.
Consider now a quantum system in one dimension, for which the 
wave function in coordinate representation is given by 
\begin{equation}
\psi_G(x)=(f_G(x:\mu, \sigma^2))^{\frac{1}{2}}, x \in \mathbb{R}.
\end{equation}
Previously, we have stated that the mean-based uncertainty relation is saturated by 
quantum wave functions whose position distributions are Gaussian. 
However, this is not true if one considers the 
semi-interquartile range as the measure of dispersion. 
The first and third quartiles for both position and momentum distributions 
can be obtained analytically in this case. 
Indeed, it is possible to numerically check that the quartiles for the position distribution are given by 
$Q_{1,G}^x = \mu -0.674 \sigma$ and $Q_{3,G}^x = \mu + 0.674 \sigma$,
so that the corresponding semi-interquartile range is $0.674 \sigma$. 
And with some algebra, the momentum distribution function, $|\phi_G(p)|^2$, yields
a semi-interquartile range of $\frac{0.337 \hbar}{\sigma}$. 
Therefore, the median-based quantum uncertainty product, 
\(\widetilde{\Delta x} \widetilde{\Delta p}\), for a quantum wave function, whose position 
distribution is Gaussian with mean \(\mu\) and standard deviation \(\sigma\), 
equals \(\frac{\hbar}{4.396}\). Note that this is much higher (precisely, 141 \%)
 than the bound obtained for the Cauchy-Lorentz position distributions. 

\subsection{Student's $t$-distribution}
\label{student}
The probability distribution function of the Student's \(t\)-distribution, due to F. R. 
Helmert, J. L{\"u}roth, ``Student'', and others, is given by
\begin{equation}
f_S(a: n)=\frac{\Gamma(\frac{n+1}{2})}{\sqrt{n \pi}~\Gamma(\frac{n}{2})} \left(1+\frac{a^2}{n}\right)^{-\frac{n+1}{2}},
a \in \mathbb{R},
\end{equation}
where $n$, which is a distribution parameter, is referred to as the number of degrees of freedom.
Here, we are interested in the case when the $t$ distribution has two degrees of freedom, 
 i.e. $n=2$. In this case, while the mean exists (and is vanishing), the 
standard deviation diverges to infinity.

Consider again a quantum system in one dimension whose wave function in coordinate 
representation is given by 
\begin{equation}
\psi_S(x)=(f_S(x: 2))^{\frac{1}{2}}=\frac{1}{(2+x^2)^{3/4}}, x \in \mathbb{R}.
\end{equation}   
The first and third quartiles corresponding to the position distribution can be 
obtained analytically, and are given by $Q_{1,S}^x=-\sqrt{\frac{2}{3}}$ and 
$Q_{3,S}^x=\sqrt{\frac{2}{3}}$, so that 
 the semi-interquartile range equals $\sqrt{\frac{2}{3}}$. 
Let us now focus on the momentum space wave function, $\phi_S(p)$ for $n=2$. 
We numerically find the quartiles corresponding to the momentum probability 
distribution by again utilizing the same
methods as for the Cauchy-Lorentz distribution, to obtain 
 $Q_{1,S}^p=-0.161 \hbar$ and $Q_{3,S}^p=0.161 \hbar$ (correct to 3 decimal figures), so that the 
corresponding semi-interquartile range is $0.161 \hbar$. 
Therefore, the median-based quantum uncertainty product, \(\widetilde{\Delta x} \widetilde{\Delta p}\),
for the quantum wave function with the Student's \(t\) distribution (for two degrees 
of freedom) as the position probability distribution, equals
\(0.131 \hbar$, i.e., \(\frac{\hbar}{7.63}\).  
The Student's \(t\) distribution is therefore somewhat midway between
Cauchy-Lorentz and Gaussian distributions
with respect to the bound on the median-based uncertainty product. 

\section{Ordering of states with respect to distance from minimum uncertainty state} 
\label{order}

We have already obtained several examples of probability
distributions which are non-Gaussian and yet provide lower
median-based quantum uncertainty products than Gaussian states.
It may seem that the Gaussian distribution still provides the minimum uncertainty state 
among distributions having finite mean and variance. To investigate this question, 
we consider the Student's t distribution with three
degrees of freedom, i.e., $n = 3$, which has a finite mean and a
finite variance. Consider therefore a quantum particle moving
in one dimension with the wave function in coordinate representation being given by
\begin{equation}
\psi_{S^\prime}(x)=(f_S(x:3))^\frac{1}{2}=\frac{\sqrt{6\sqrt{3}}}{\sqrt{3}(3+x^2)}, x \in \mathbb{R}
\end{equation}
The first quartile, $Q^x_{1,S^\prime}$, of the position distribution, is given
by $\sin {\theta} + \theta +\frac{\pi}{2} = 0$, where $\theta = \frac{1}{2}\tan ^{-1}
\frac{Q^x_{1,S^\prime}}{\sqrt{3}}$, leading to $Q^x_{1,S^\prime}= -0.765$, correct 
to three significant figures. By symmetry, $Q^x_{3,S^\prime}= 0.765$, so that the
semi-interquartile range is $0.765$. In the momentum space, we have found the quartiles
numerically. They are $Q^p_{1,S^\prime}= -0.200\hbar$ and $Q^p_{3,S^\prime}= 0.200\hbar$
(correct to 3 decimal figures), so that the corresponding semi-interquartile range is 
$0.200\hbar$ . Therefore, for $n = 3$, the median-based quantum uncertainty product,
$\widetilde{\Delta x} \widetilde{\Delta p}$, is equal to $0.153\hbar$ , i.e., 
$\frac{\hbar}{6.54}$. The Student's
t distribution with three degrees of freedom is therefore a non-Gaussian distribution with
finite mean and finite variance, and yet better off in the race for the 
minimum uncertainty state than the Gaussian.
This therefore implies that the ordering of states with respect to distance from the minimum 
uncertainty state is altered in the case of the median-based uncertainty relation, in 
comparison to the mean-based one.

The F distribution, due to R. Fisher and G. A. Snedecor,
does not have finite mean and variance for the degrees of freedom
$d_1 = 5$ and $d_2 = 2$, and is a probability distribution that is
asymmetric around its median. For a quantum particle 
moving in one dimension and whose wave function is the square
root of such an F distribution, the median-based uncertainty
product provides a value much higher than that of the Gaussian distributions.


\section{Completeness of polynomials and uncertainty relation}
\label{completeUR}

We now invoke the completeness~\cite{sobi-power-law} of Hermite polynomials in \(L_2(-\infty, +\infty)\)
to determine the minimum uncertainty state 
among quantum states corresponding to systems of a single particle moving in one dimension.
We Haar uniformly generate such functions numerically, obtaining convergence 
by considering polynomials until degree 8, and find that 
the median-based uncertainty relation can be expressed as 
\( \widetilde{\Delta x} \widetilde{\Delta p} \geq \frac{\hbar}{5.88}\).
The minimal state is not Gaussian, as all Gaussian states provide a value of 
\(\frac{\hbar}{4.396}\) for the median-based uncertainty product.
%
%
This is to be compared with the mean-based quantum uncertainty product 
(see relation (\ref{ur_xp-eiTa-purono}), where \(r=2\)), which is saturated by the 
Gaussian position distributions. Note that the numerical search have 
disregarded the Cauchy-Lorentz and Student's $t$ distributions, possibly relegating them to some zero-measure sets, as both provide a lower minimum for the median-based uncertainty product. Therefore, 
among the distributions considered in this paper, 
we obtain 
\begin{equation}
 \widetilde{\Delta x} \widetilde{\Delta p} \geq \frac{\hbar}{\tilde{r}},
\end{equation}
where \(\tilde{r}=10.6\).

Physical quantities like entropy and entanglement~\cite{horo-rmp} are extensive properties of a system, and typically scale with the size of the system. Therefore they diverge when one considers Haar uniform generation of states in the entire state space. It is crucially important to analyze such extensive physical quantities under physically motivated resource constraints, like those on energy~\cite{yuen70, caves66, holevo59, holevo63, lloyd90, Giovannetti91, Giovannetti70, gio70, gio92, sen95, sen75, wolf98, Serafini40, lupo104, lupo53, gio8}. In this paper, we have generated superpositions of Hermite polynomials Haar uniformly, and find that the product of semi-interquartile ranges of the position and momentum converges with increasing degree of the Hermite polynomials.
Similar convergence is seen for the product of variances of position and momentum, which we have also checked within the scenario of numerical Haar uniform generation of Hermite polynomials of increasing degrees (and found that considering Hermite polynomials until degree 5 provides convergence to \(1/2\) up to 6 decimal points in units of \(\hbar\)). Such products of spreads are therefore intensive physical quantities, and 
even though an energy-like constraint is not absolutely necessary in these cases, it is certainly interesting to analyze intensive quantities under a resource constraint.

Although we have restricted ourselves in this paper to pure quantum states of a system in one dimension, 
the considerations can be extended to mixed states and higher dimensions.

Additionally, it is important to mention here that the considerations can be taken over to quantum systems with discrete degrees of freedom. However, since spreads of observables in discrete systems are always finite, it is natural to consider sums of spreads (or squares thereof) for analyzing uncertainty relations in such systems.
As an example,  let us consider the two observables, $\sigma_x$ and $\sigma_y$, the Pauli matrices of a spin-1/2 system.
We now show that for an arbitrary quantum spin-1/2 state $|\psi\rangle$,
$ \widetilde{\Delta \sigma_x}^2 + \widetilde{\Delta \sigma_y}^2 \neq 0.$
 Clearly, $\widetilde{\Delta \sigma_x}$  and  $\widetilde{\Delta \sigma_y}$ can be either 0 or 1. Therefore, the sum of 
 their squares can only be among $0, 1, 2$. We, however, show that quantum mechanics dictates that $\widetilde{\Delta \sigma_x}^2 + \widetilde{\Delta \sigma_y}^2 \geq 1$. 
 To prove this, let us assume that $\widetilde{\Delta \sigma_x}=0$, which implies that both the first and the third quartiles
 are equal to the \(+1\) eigenvalue of \(\sigma_x\) or both are equal to the \(-1\) eigenvalue of the same.
In such a case, 
the state $|\psi\rangle$ can be expressed as 
\begin{equation}
|\psi\rangle = \sqrt{p}~|-x\rangle + e^{i \theta}  \sqrt{1-p}~|+x\rangle,
\end{equation}
with the condition $3/4 < p \leq 1$. \(\theta\) is a real number in \([0,2\pi)\), and \(|\pm x\rangle\) are  the eigenstates of $\sigma_x$.   
Now, 
\begin{equation}
|\langle \pm y|\psi\rangle|^2=  \frac{1}{2} \big[1\pm 2 \sqrt{p (1-p)}\sin \theta \big],
\end{equation}
where $|\pm y\rangle=\frac{1}{\sqrt{2}} (|0\rangle \pm i |1\rangle)$ are the eigenstates of $\sigma_y$.
For $\widetilde{\Delta \sigma_y}$ to be zero, we must have either of \(|\langle \pm y|\psi\rangle|^2\) greater than 3/4, and the latter is disallowed, because \(p>3/4\). 
This shows that if  $\widetilde{\Delta \sigma_x}=0$, then we must have $\widetilde{\Delta \sigma_y} \neq 0$, for arbitrary quantum states of a qubit. A similar argument holds with the roles of \(\sigma_x\) and \(\sigma_y\) reversed, implying that $\widetilde{\Delta \sigma_x}^2 + \widetilde{\Delta \sigma_y}^2 \geq 1$. 

\section{Conclusions}
\label{conclude}

In this paper, we conceptualized a quantum uncertainty relation for arbitrary quantum states that 
has a wider applicability than the traditional one. The traditional one is based on the concept of the 
mean -- it is a bound on a product of standard deviations, a key role in whose definition is played by the mean. 
%
By contrast, the uncertainty relation 
presented here is based on the concept of the median, and its corresponding dispersion quantity, namely the semi-interquartile range. 
There are distinct situations where the mean of a distribution does 
not provide the best representative value of the distribution, including situations where the mean 
or the standard 
deviation does not exist. 

In the course of working with the mean, it was realized that quantum wave functions whose position distributions are 
Gaussian, are the minimum-uncertainty states.
 The picture changes when we deal with the median, and the Cauchy-Lorentz position distributions seem to be the 
functions of choice for being the same.
Moreover, 
the ordering of quantum states with respect to their distance from the 
minimum uncertainty state is also drastically altered by moving over to a measure of 
spread with universal applicability.

\section{Acknowledgment}
We thank Paul Busch for helpful comments. A.B. acknowledges the support of the Department of Science and Technology (DST), Government of India, through the award of an INSPIRE fellowship.

\end{document}